\documentstyle[aps,prl,psfig]{revtex}
\begin{document}
\draft \twocolumn
[\hsize\textwidth\columnwidth\hsize\csname@twocolumnfalse\endcsname
\title{Solar Flares as  Cascades of Reconnecting Magnetic Loops} \author{D. Hughes,$^1$ M. Paczuski,$^1$ R.O. Dendy,$^2$ P. Helander,$^2$ and K.G. McClements$^2$}

\address{$^1$ Department of Mathematics, Imperial College of Science, Technology, and Medicine, London, SW7 2BZ, UK\\ $^2$ 
UKAEA, Culham Science Centre, Abingdon, Oxfordshire OX14 3DB, UK}

\date{\today}

\maketitle

\begin{abstract}

A model for the solar coronal magnetic field is proposed where
multiple directed loops evolve in space and time.  Loops injected
at small scales are anchored by footpoints of opposite polarity
moving randomly on a surface.  Nearby footpoints of the same
polarity aggregate, and loops can reconnect when they collide.
This may trigger a cascade of further reconnection, representing a
solar flare.  Numerical simulations show that a power law
distribution of flare energies emerges, associated with a scale
free network of loops, indicating self-organized criticality.

\end{abstract}

]

\narrowtext

One can think of the magnetic field beneath the photosphere of the sun
as a giant ball of yarn.  Magnetic flux tubes are
irregularly expelled into the corona as loops anchored to
the photosphere.  They form the building blocks for larger scale
magnetic structures such as  active regions or the magnetic
carpet.  A solar flare is a rapid change in a strong, complicated
coronal magnetic field\cite{Zirker}. 
According to a physical picture proposed by
Parker, turbulent plasma flow below the photospheric surface drives
the anchored flux tubes into complex, stressed
configurations\cite{parker}. When local magnetic field gradients
become sufficiently steep, a plasma instability allows the coronal
magnetic field to change its topology via reconnection, suddenly
releasing energy\cite{mhd}.

The observed statistics of peak X-ray flux distributions\cite{Dennis},
the energy released\cite{Aschwanden}, and the quiescent time intervals
between solar flares\cite{Boffetta,wheatland} are all characterized by
power law distributions. The energy distribution is
particularly striking, exhibiting scale-free behavior over more than
eight decades in energy\cite{Aschwanden}. These statistics
indicate\cite{LH} that the solar corona may be in a state of
self-organized criticality (SOC)\cite{btw}. Such a view implies that large
classical X-ray flares and small transient brightenings in the EUV spectrum
 are not fundamentally
different; they are large and small avalanches of
reconnection events. In fact, the statistics of solar
coronal behavior shares some common features with other intermittent
scale free phenomena such as earthquakes, forest fires, traffic,
evolution,  and turbulence\cite{bak,turcotte}.

Here we introduce a dynamical model of multiple magnetic loops that
are randomly driven at their footpoints, and can interact.  A pair of
footpoints, of opposite polarity, anchors each directed loop to a two
dimensional surface, representing the photosphere.  Footpoints of the
same polarity interact by local aggregation on the surface. Loops also
interact by exchanging footpoints when they collide.  A single
rewiring, or reconnection, can lead to more loop collisions and
thereby trigger a cascade of further reconnection. These cascades of
magnetic loop reconnection are identified with solar flares.  

Loop configurations observed in numerical studies of the model (see
Fig. 1) are qualitatively similar to the magnetic carpet deduced from
observations of the quiet sun\cite{Zirker,mhd,website}.  The pattern
of loops dynamically forms a scale free network\cite{barabasi,strogatz},
where the number of loops emerging from each
footpoint is distributed as a power law.  This network gives rise to a
power law distribution of flare energies.  Results from our model also
suggest that the probability distribution of net magnetic flux emerging from
small areas of the photosphere exhibits power law behavior, while the
distribution of lengths of flux tubes in the corona is exponential.
Our results support the idea that solar flare statistics reflect
coronal SOC resulting from magnetic reconnection.

\begin{figure}[bt] \centerline{\psfig{figure=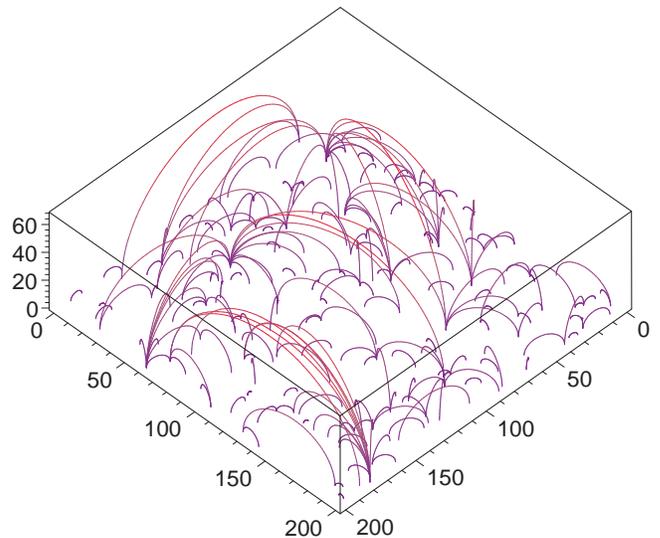,width=\columnwidth}}
\vspace{0.25cm} \caption{Snap shot of a
configuration of loops in the steady state with $L=200$ and
$m=1$ (see text). } \label{ps1} \end{figure}

The conventional techniques of plasma physics\cite{mhd} have
generated substantial progress in the theoretical understanding
and computational modelling of local reconnection in laboratory,
space, and solar plasmas.  However, it remains computationally
prohibitive to use the equations of plasma physics to describe
extended spatial regions containing multiple evolving flux tubes that
interact via magnetic reconnection. This has motivated the
adoption of simpler numerical approaches.

Lu and Hamilton (LH) used  a cellular automaton as a  simple
model\cite{LH} for the reconnection process leading to flares. The
LH avalanche model is similar to the original sandpile model of
Bak, Tang, and Weisenfeld\cite{btw}.  It is conceptually far
reaching, and has stimulated many further investigations (see for
example\cite{review,otherlh}), as well as other 
cellular automata models\cite{zirker-cleveland}.

The multiloop reconnection model introduced here is not a cellular automaton
 and differs fundamentally from previous approaches.  But
it makes explicit contact with the physics of solar flares in two
ways.  First, it retains the basic topological constraints that
the high conductivity of the corona and the physics of
reconnection impose on magnetic field evolution.  Second,  it
incorporates turbulent footpoint motion and the resultant driving
of coronal magnetic field energy. 
 The key elements of our model
are as follows:

1.\ \ {\it Loop structure.} A loop 
labels the midline of a magnetic flux tube, which  has a diameter
of order $100
km$ at its footpoint on the photosphere\cite{parker}.
Therefore, the coronal magnetic field is
represented by numerous infinitesimally thin directed loops, which
are semicircles emerging from the $(xy)$ plane.   Each loop has a
positive footpoint, where magnetic flux emerges from the
photosphere, and a negative one, where flux returns.  The size of
the system in the $(xy)$ plane, which represents a region of the
photospheric surface, is $L \times L$.  Loops are labeled by an
integer, $n$, and the positions of the two footpoints of the $n$th
loop are labeled in the $(xy)$ plane by ${\bf r}^+_n$ and ${\bf
r}^-_n$. Each loop is associated with one unit of magnetic flux, so that the
magnetic energy of a loop is proportional to its length $l_n= {\pi\over 2}
|{\bf r}^+_n -{\bf r}^-_n|$.  The sum of the lengths of the loops,
$E = \sum_n l_n$, is then a measure of the total magnetic energy
of the system, and changes in the value of $E$ correspond to the
magnetic energy release in solar flares.

2.\ \ {\it Footpoint motion.} Footpoints are considered to be
passively convected by the turbulent plasma motion beneath the
photospheric surface\cite{parker}. This is represented in the present
model by a random walk. At an update step, an arbitrary footpoint
is chosen at random and its position is moved, ${\bf r}
\rightarrow {\bf r} + \Delta {\bf r}$. The vector $\Delta{\bf r}$
has length and angle chosen randomly from uniform distributions
between 0 and 1, and 0 and $2\pi$, respectively.  If the initial
loop lengths are small, random footpoint motion will tend on average to
increase the length of the loops. In this way, photospheric
turbulence pumps magnetic energy into the coronal magnetic field.
Note that footpoints can have more than one loop attached, as per
step 5.

3.\ \ {\it Loop injection and submergence.} The coronal magnetic
field is also driven by injection of new loops from beneath the
photospheric surface\cite{Zirker,parker}. In
the model, small loops are injected into the system at random
locations, with footpoints initially separated by a distance
$l_{nl}=4$. Loops with footpoints closer than distance $l_{min}=2$
are removed from the system.  The precise length scales of these
two processes do not effect the critical properties of the system.
The essential feature is that at small length scales the model
dynamically maintains a flow of loops. Thus the magnetic field of
the corona is represented as an open system.

\begin{figure}[bt] 
\centerline{\psfig{figure=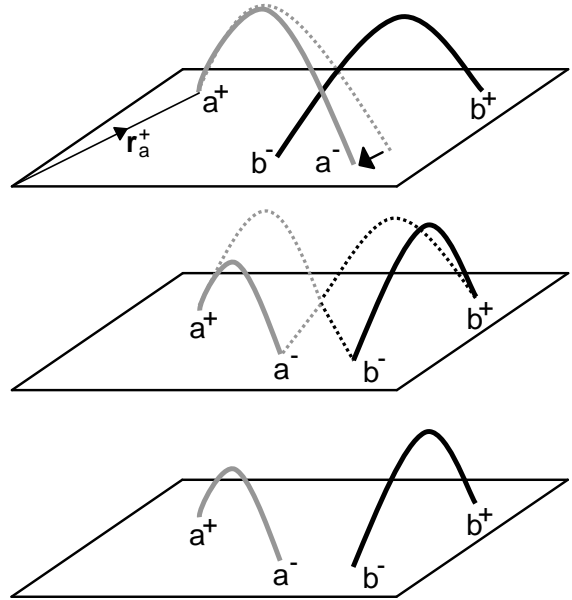,width=\columnwidth,clip=}}
\vspace{0.25cm} \caption{Diagram showing the process of a
reconnection event, from top to bottom. In frame 1 loop $a$ moves
from its previous position (dashed line) and crosses loop $b$. In
frame 2 the loops exchange footpoints and move to their rewired
state. Frame 3 shows the final relaxed configuration.} \label{ps2}
\end{figure}

4. \ \ {\it Loop reconnection.}  Flux freezing in the coronal
plasma is relaxed when local field gradients become sufficiently
steep: the magnetic field then changes its topology, and releases
energy by reconnection events. Reconnection can occur in our
model when either (a) two loops collide in three dimensional space, or
(b) two footpoints annihilate as explained in step 6.  In the first
case, the midlines of two flux tubes have crossed resulting in a
strong magnetic field gradient at that point. The flux emerging from
the positive footpoint of one of the reconnecting loops is then no
longer constrained to end up at the other footpoint of the same loop,
but may instead go to the negative footpoint of the other loop (see
Fig. 2).  Reconnection is only allowed if it shortens the combined
length of the two colliding loops.  This process is rapid compared to
the driving, because information is transmitted along the 
flux tubes at the
Alfv\'{e}n speed.

If rewiring occurs, it may happen that one or both  loops
need to cross some other loop in order to reach its  rewired
state. Thus  a single reconnection between a pair of loops can trigger
 an avalanche-like cascade of causally related reconnection
events. The reconnection dynamics of multiply connected footpoints
that we used in the numerical simulations  is a straightforward
extension\cite{details}.

5. \ \ {\it Footpoint aggregation.}  It has been observed that
footpoints of the same polarity can merge when they
approach\cite{parker}.  In addition, calculations of the coronal
magnetic field, based on measurements of the photospheric field,
indicate that flux tubes starting at a particular footpoint can
terminate at several different footpoints of opposite
polarity\cite{Zirker,mhd,website}.  In our model, if a footpoint moves
within the distance $l_{min}$ of another footpoint of the same
polarity, it is reassigned to the latter footpoint's position and they
move as one footpoint thereafter.

6. \ \ {\it Footpoint annihilation.}  Footpoints of opposite polarity,
belonging to different flux tubes, can annihilate when they approach
on the photosphere\cite{mhd}.  When footpoints of opposite polarity
belonging to different loops approach within a distance $l_{min}$ in
the model, both footpoints are eliminated and the remaining two
footpoints are attached, forming one loop, where before there were
two.  The annihilation dynamics
of footpoints with more than one attached loop can be implemented in
different ways\cite{details}.  The
scaling behavior of the model does not appear to be sensitive to the precise
algorithm.  Footpoint annihilation may cause collisions between
loops leading to further reconnection, as per step 4.

7. \ \ {\it Time stepping and boundary conditions.}  In each time
step  of the model, either a footpoint is chosen for random motion
or a new  loop is added.  This is governed by a control parameter,
the stirring  rate $m$, such that the number of footpoint updates
that separate an  update step where a new loop is injected into
the system is $m$ times  the number of footpoints in the system at
that time.  We study a
system with open reflective  boundary conditions; if a footpoint
attempts to move outside the $L  \times L$ box in the $(xy)$
plane, it is elastically reflected back  into the box.

The configuration of  loops slowly evolves in response to
the driving, aggregation and reconnection processes described
above.  It reaches a dynamic equilibrium whose statistical
character is independent of the initial conditions.  Below,
numerical simulation results are presented for system size $L=200$
with a range of stirring rates $m$ from 0.01 to 1, including
approximately $10^7$ avalanches of reconnection events in the
steady state. The observed power law behavior is robust on varying
the stirring rate, even though the total magnetic
energy and the fluctuations in the total energy vary widely.

Figure 3 shows the probability distribution of the number $k$ of
loops emerging from each footpoint, $P_{foot}(k)$. It is
well described by a power law $P_{foot}(k) \sim k^{-\gamma}$, with
$\gamma =2.0 \pm 0.1$ over more than two decades in number of links.  The
cutoff in the distribution increases with decreasing $m$ for fixed
$L$. This is due to the fact that the total number of loops in the
system is increasing.

\begin{figure}[bt] \centerline{\psfig{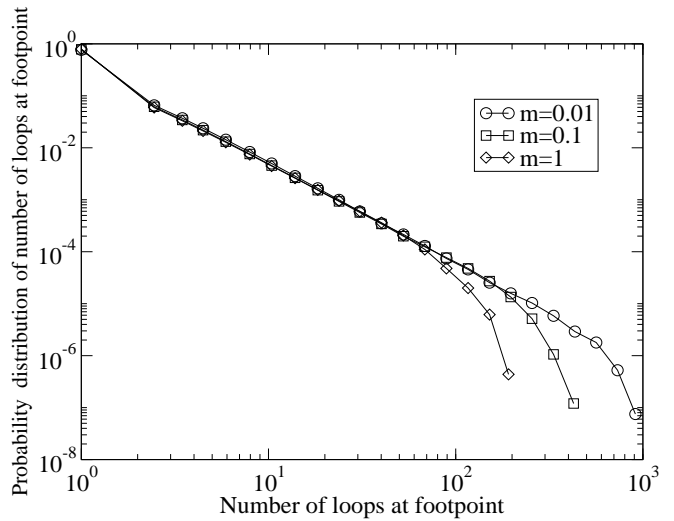}}
\vspace{0.25cm} \caption{Distribution of number of loops (links)
per footpoint (node) . The slope on the log-log scale is -2.
Thus $P_{foot}(k) \sim k^{-\gamma}$ with $\gamma=2$.}
\label{ps3} \end{figure}

Thus, the system of loops that emerges forms a scale
free network, with a cutoff determined only by the number of loops
(links) in the system. Scale free networks, in the form of graphs,
have been invoked recently to describe a wide variety of phenomena
including e.g.  citation networks, the internet, and some
biological systems; for reviews see Refs.~\cite{barabasi,strogatz}.  
Unlike most of these examples, our
network is embedded in three-dimensional space, and geometrical
constraints are crucial to the dynamical self-organization by
which this network emerges.

Figure 4 shows the probability distribution for flare event energy
release, $P_{flare}(E)$. This also follows a power law
$P_{flare}(E) \sim E^{-\alpha_E}$, with $\alpha_E =3.0 \pm 0.2$.  This value
is much larger than the exponent obtained from the LH type avalanche
models, which give $\alpha_E \simeq 1.4$\cite{LH}.  The
distribution of individual loop lengths in our multiloop model,
unlike that of flare energies, is exponential.

The total energy released in a solar flare is not directly
measured, but  is inferred from narrow frequency band observations
of photon fluxes. A number of simplifying assumptions are used to
relate these fluxes to total energy release, as explained in e.g.
Refs.~\cite{review,norman,geometry}. There is some
uncertainty in the value of the exponent of the measured energy
distribution of solar flares, with reported values ranging from
$\alpha_E = 1.5 {\rm \ \ to \ \ } 2.6$\cite{review,norman}.
However,
if a consistent set of geometrical assumptions are made, the upper
value may be reduced from 2.6 to approximately 2.1\cite{geometry}.  
Our present result is outside  this range, but is nevertheless
consistent with Parker's conjecture
for coronal heating\cite{parker} by nanoflares which minimally requires that
$\alpha_E \geq 2$\cite{Zirker,Aschwanden,review}.

Another, possibly related
 difference from observations concerns temporal
correlations.  Solar flares, like earthquakes and other
intermittent phenomena, exhibit power law statistics for the
waiting times between events whose magnitude
exceeds some threshold\cite{Boffetta,wheatland,bak1}. The model presented
here, like LH avalanche models, does not exhibit any 
temporal correlations in the waiting times between events.  This
may be due to the assumption
of complete randomness in the external driving.

\begin{figure}[bt] 
\centerline{\psfig{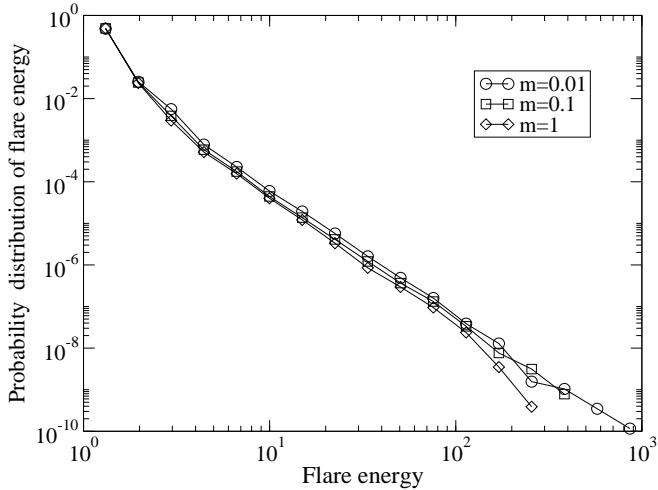}}
\vspace{0.25cm} \caption{Distribution of flare energies in the
multiloop reconnection model.  The slope on the log-log scale is  -3. 
}
\label{ps4} \end{figure}

Flux tubes are tangled into complex configurations by the
turbulent plasma below the photosphere\cite{parker}.  As a starting
point, we have modeled the turbulent forces as inducing both
completely random footpoint motion and completely random injection of
new loops. These assumptions neglect some features of the
photosphere that involve spatiotemporal correlations, such as
granulation of the surface, phenomena such as homologous flares, and
the observation that active regions, where most large flares occur,
have very high rates of magnetic flux emergence compared to the quiet
sun\cite{Zirker}. Since we describe the magnetic loop
configurations as being passively driven by an external source, our
model could easily be extended to incorporate specific spatiotemporal
correlations in the driving, in particular in the rate that loops are
added to the system at different places.

Following a suggestion by Wheatland\cite{wheatland} 
that nonstationary driving could induce waiting time
correlations, Norman {\it et al} have studied LH avalanche models with
a driving rate modulated by a random walk and found both a change from
exponential to a power law distribution of waiting times, as well as
a small change in the critical index $\alpha_E$\cite{norman}.  Perhaps a
similar nonstationary driver would have comparable effects in our model.

We have introduced a model of the solar coronal magnetic field
consisting of multiple reconnecting magnetic loops.  We believe that
this model captures essential elements of the physics governing such
structures. Our approach applies to situations involving more loops,
and over greater length and time scales, than are accessible to
traditional analytical and computational techniques based on the
underlying equations.

 The striking feature of our multiloop reconnection model is the
dynamical self-organization of a magnetic field which gives rise to a
power law distribution of solar flare energies, and which forms a
scale free network that qualitatively resembles the actual coronal
magnetic field.  The behavior, $P_{foot}(k)\sim
k^{-\gamma}$, could perhaps be tested by imposing a fine grid on the
photosphere, depending on resolution capabilities, and measuring the
probability distribution of total magnetic flux within each grid cell.
Our model results suggest that this distribution would have power law
behavior, with index $\gamma$, while the distribution of lengths of
flux tubes would be exponential.

We thank P. Cargill and S.C. Cowley for discussions.
This work was funded in part by the UK Department of Trade \&
Industry and the EPSRC.

\end{document}